\documentclass[aps,prb,superscriptaddress,twocolumn,floatfix]{revtex4-1}
\usepackage[permil]{overpic}
\usepackage{graphicx}
\usepackage{latexsym}
\usepackage{amsmath}
\usepackage{amssymb}
\usepackage{upgreek}
\usepackage{xcolor}
\usepackage[colorlinks]{hyperref}
\hypersetup{allcolors=blue}

\begin{document}

\title{Planar defects as a way to account for explicit anharmonicity in high temperature thermodynamic properties of silicon.}
\author{M. V. Kondrin}
\email{mkondrin@hppi.troitsk.ru}
\affiliation{Institute for High Pressure Physics RAS, 108840 Troitsk, Moscow, Russia}
\author{Y.B. Lebed}
\affiliation{Institute for Nuclear Research RAS,  Moscow, Russia}
\author{V.V. Brazhkin}
\affiliation{Institute for High Pressure Physics RAS, 108840 Troitsk, Moscow, Russia}

\begin{abstract}
Silicon  is  indispensable  in semiconductor industry. Understanding its high-temperature thermodynamic properties is essential  both for theory and applications. However,  first-principle description of high-temperature thermodynamic properties of silicon  (thermal expansion coefficient and specific heat) is still  incomplete. Strong deviation of its specific heat at high temperatures from the Dulong-Petit law suggests substantial contribution of anharmonicity effects. We demonstrate, that anharmonicity is mostly due to two transverse phonon modes, propagating in (111) and (100) directions, and can be quantitatively described with formation of the certain type of nanostructured planar defects of the  crystal structure. Calculation of these defects' formation energy  enabled us to determine their input into the specific heat and thermal expansion coefficient. This contribution turns out to be significantly greater than the one calculated in  quasi-harmonic approximation.  
\end{abstract}

\maketitle

\section{Introduction}

Silicon is a widely used semiconductor with immense impact in industry. Understanding its high-temperature thermodynamic properties is necessary both for theory and applications. Also, due to its archetypal diamond structure,  silicon is among the touchstones for testing new theories. Still,  there are black spots in its first principle description, in particular,  high-temperature thermodynamic properties. In large extent this is due to the fact that high-temperature properties are connected to anharmonicity effects, which are difficult to describe by ab-initio methods alone. 

In fully harmonic approximation both isochoric specific heat $c_V$ and isobaric thermal expansion coefficient $\alpha_V$ in solids at the limit of high temperatures both should converge to constant value \cite{landafshitz:v}. For $c_V$ this is a result of the classical Dulong-Petit law. In the latter case the constant value of $\alpha_V$ at high temperature is the result of quasiharmonic approximation where free energy taken  as the sum of potential energy of atoms assumed to be confined by the harmonic potential and phonons is minimized at certain temperature with respect to volume. This results in equation \cite{landafshitz:v}:

\begin{equation}
\alpha_V=\frac{\gamma c_V}{B_0 V}
\label{eq:bv}
\end{equation} 

Here $\gamma=-\frac{V}{\Theta}\frac{d\Theta}{dV}$ is the Gr\"{u}neisen parameter which in zeroth approximation is a constant independent of temperature, $V$ is specific volume and $B_0=-V\frac{dP}{dV}$ is a bulk modulus at zero pressure. $\Theta$ in the definition of the Gr\"{u}neisen parameter can be taken as characteristic frequency of phonon spectrum like, for example, the Debye temperature. Still, this is an oversimplification, because $\gamma$ can be different for different phonon modes (even negative for transverse acoustic  modes in silicon and germanium which lead to negative thermal expansion in these crystals at low enough temperature \cite{swenson:jpcrd83}), therefore parameter $\gamma$ acquires dependence on temperature. Nonetheless, at temperatures above the frequency of the highest phonon mode $\gamma$ tends to constant value and, so, the expansion coefficient does the same.

In quasiharmonic approximation $c_V$ at high temperature does approach 3$R$ level and the specific heat at constant pressure $c_P$ (which in contrast to $c_V$ is usually measured in experiments) is connected to $c_V$ by thermodynamic identity:

\begin{equation}
c_P-c_V=\alpha^2_VB_0VT
\end{equation}
, where $\alpha_V$ -- volume thermal expansion coefficient, $V$ -- specific volume , $T$ - temperature.

In solids, due to comparatively low value of $\alpha_V$, the difference between $c_P$ and $c_V$ at melting temperature usually does not exceed few percents and therefore can be neglected. However, in real solid materials both high temperature expansion coefficients and isochoric heat capacity deviate from constants,  as a result of explicit anharmonicity effects\cite{born:zp21,wallace:pr65,cowley:rpp68}. Precise values of this deviations in silicon, prescribed by quasiharmonic approximation,  are not strictly known, because in various measurements of high temperature specific heat \cite{gerlich:jap65,desai:jpcrd86,yamaguchi:jtac02} and expansion coefficients \cite{swenson:jpcrd83,maissel:jap60,watanabe:ijt04,dutta:pssb62,okada:jap84,roberts:jpdap81}  differ from one another by about 10 \%, but all of them are definitely not constant. Possible origin of this scatter we will discuss below.
 
{\em Ab-initio} computation of anharmonic effects at finite temperatures requires either resource-demanding {\em ab-initio} molecular dynamics simulations (see e.g. the computation of thermodynamic parameters of aluminum \cite{grabowski:prb09}) or, under some simplifying assumptions which let to sample potential energy  surface of material in vicinity of equilibrium point, by strictly DFT methods at zero temperature (so called stochastic-initialized temperature-dependent effective potential/s-TDEP method \cite{kim:pnas18}). The latter approach was recently applied to describe thermal expansion coefficient of  silicon \cite{kim:pnas18}. Still, in this case we do not agree with authors remark that ``a simple physical model for the anomalous thermal expansion of silicon is unlikely because different effects contribute to the thermal expansion. In particular, anharmonicity and nuclear quantum effects are difficult to formulate as a simple 3D model''\cite{kim:pnas18}. In fact, we are able to demonstrate, that major part of high temperature anharmonicity in silicon is connected to its' few (namely two) anharmonic modes,  which lead to formation of metastable state (planar defect) with relatively low energy.

This analogy between defect and anharmonic phonon oscillation was proposed in the number of our recent works \cite{kondrin:drm20,kondrin:prl21,kondrin:pssb22} and was successfully used to resolve discrepancies in experimental thermodynamic properties of graphite near the melting temperature  \cite{savva:pu20,savva:c16,kondratyev:prl19,kondrin:prl21}. In the nutshell, this analogy relates strong anharmonicity to finite displacement of atom layers caused by shear phonon modes, and subsequent reconstruction of interatomic bonds. It is this reconstruction which leads to drastic lowering of potential energy at certain values of atom displacement, and can be effectively described as anharmonicity of interatomic potential. On the other hand, reconstruction of interatomic bonds can be regarded as extended planar defects, and its thermodynamic properties can be calculated using formation energy value of this defect. This approach has its strong sides because it can be developed by zero temperature ab-initio theory and requires knowledge of only few lowest energy defects which mostly contribute to thermodynamic properties of material. We will demonstrate, that in silicon only two phonon modes have to be taken into account to successfully describe its high-temperature thermodynamic properties (both thermal expansion and specific heat). 

\section{Explicit anharmonicity in solids}

It should be noted, that in general, the expansion coefficient of silicon is regarded to be small at the temperatures below 1000 K, so silicon was proposed as the expansion coefficient standard \cite{swenson:jpcrd83,watanabe:ijt04}. Nonetheless, at high temperatures it increases due to explicit anharmonicity \cite{born:zp21,wallace:pr65,cowley:rpp68}. Considering anharmonicity effects it is appropriate to discern two types of anharmonicity -- a uniform one (which we call bulk anharmonicity), and the transverse one, which is due to bond reconstruction, caused by transverse phonon modes mentioned above. Bulk anharmonicity is caused by softening of elastic constant at high temperatures. Although, in some works this contribution is called quasiharmonic, but it  heavily depends on anharmonicity of elastic modulus on pressure(volume) change, so it includes purely anharmonic input. This effect in silicon is evident from experimental measurements of Young's moduli along different directions \cite{vanhellemont:ecst14} and speed of sound at various directions \cite{goncharova:ftt83en}. Still, in quasiharmonic approximation this effect can be described by substitution of $B_0V$ term in Eq.~(\ref{eq:bv}) by the temperature dependent analogue. So, bulk modulus at finite temperatures can be replaced by the relation $B=B_0+B'\Delta P=B_0-B'B_0 \Delta V/V= B_0-B'B_0\int\limits_0^T \alpha_VdT$. Here, $B'$ is the pressure derivative of bulk modulus which can also be obtained by zero temperature {\em ab-initio} calculations. Obviously, in this approximation $V$ should be replaced by $V+V\int\limits_0^T\alpha_v^0(T)dT$. So, the contribution to thermal expansion coefficient due to isotropic softening of bulk modulus can be handled by iterative procedure, that is the next iteration of expansion coefficient can be written as ($\alpha_V^0$ is just the quasiharmonic approximation described by Eq.~\ref{eq:bv}):
\begin{equation}
\alpha_V^1=\alpha_V^0(1+(B'-1)\int\limits_0^T\alpha_v^0(T)dT)
\label{eq:a1}
\end{equation} 
The difference between bulk and transverse anharmonicity can be related to the difference between quasiharmonic and explicit anharmonicity. The latter is supposed to lead to the thermal dependence of phonon modes at constant volume. Still we should point out that the transverse anharmonicity leads also to another manifestation namely to the random splitting of phonon modes, leading to broadening of phonon dispersion curves. This effect is observed in experiment \cite{kim:prb15}.

\begin{figure}
\includegraphics[width=\columnwidth]{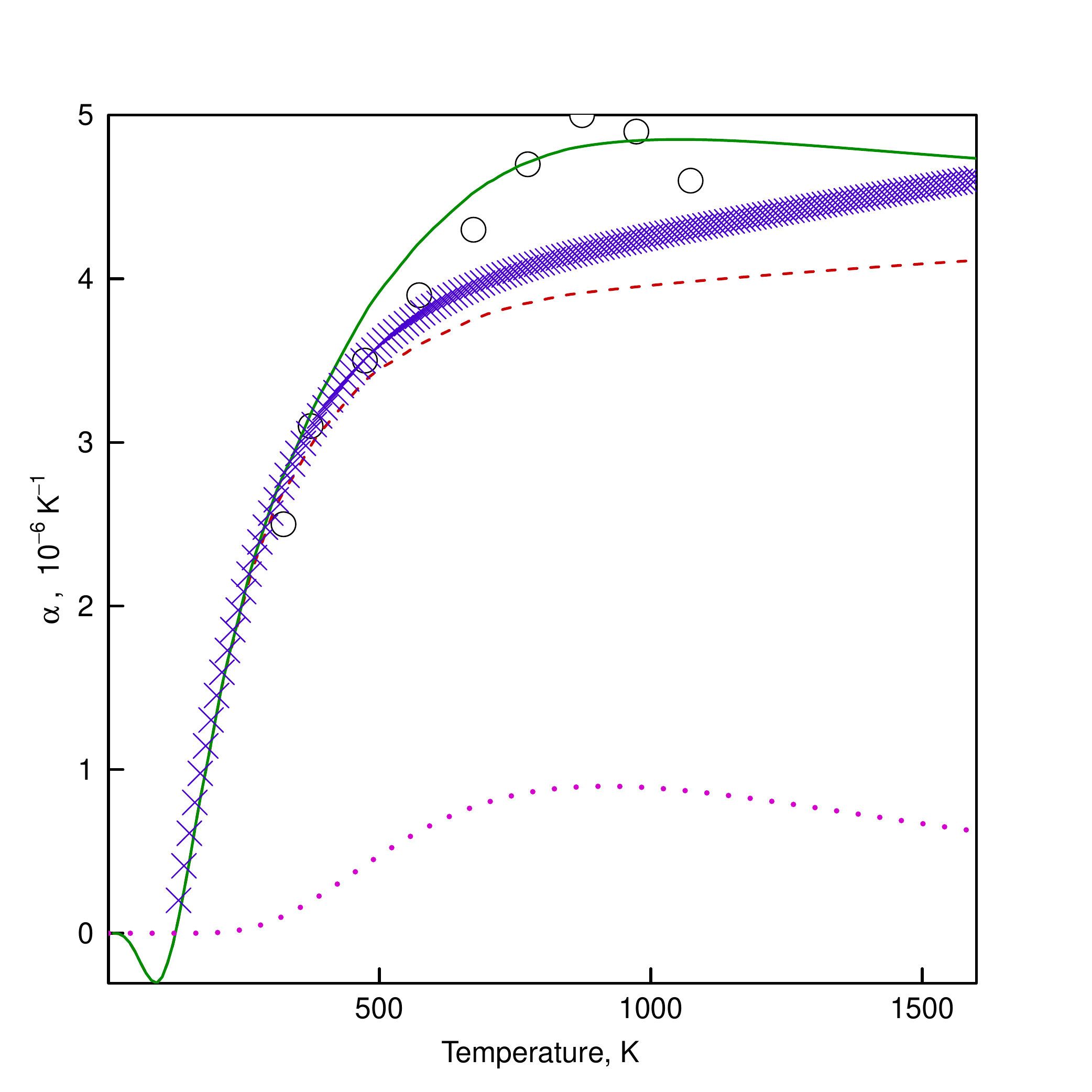}
\caption{Various theoretical and experimental data of linear expansion coefficient of silicon. Red dashed curve -- quasiharmonic approximation uniform expansion coefficient given by Eq:~\ref{eq:a1}. $\times$ -- experimental data by Okada {\em et. al.} \cite{okada:jap84} measured by X-Ray technique. $\circ$ -- experimental data of Maissel \cite{maissel:jap60} measured by optical technique. Magenta dotted curve -- theoretical contribution from (111) planar defects. Green solid curve -- sum of two inputs to linear expansion coefficient from uniform dilatation and expansion caused by (111) planar defects. Note that the last curve is not a fit of experimental data because it contains no fitting parameters bat rather a comparison of theoretical and experimental results.}
\label{f:1}
\end{figure}

If we take the value of the temperature dependent  Gr\"{u}neisen parameter from {\em ab-initio} calculation for pure silicon done before \cite{wei:prb94a}, and values of $B' \approx 4$  and $B_0 \approx 90$ \cite{wang:zfna15} the value of $\alpha^1_V(T)$ will not deviate much from constant at high temperature up to the melting point (see Fig.~\ref{f:1}). Its value  is even lower than the expansion coefficient obtained by X-Ray measurements by Okada {\em et al.} \cite{okada:jap84}. Therefore, we should seek for another contribution to the expansion coefficient of silicon. We argue, that this previously unaccounted for contribution to thermal expansion coefficient of silicon can be described by taking into consideration only two planar defects.

\begin{figure*}
\caption{Silicon (111) (panel a) and (100) (panel b) defect structures. Atoms of pristine silicon slabs are shown by solid green, and defect layer atoms - by translucent color. For (111) structure the two different interface layers are shown by different color.}
\begin{overpic}[width = 0.49\textwidth]{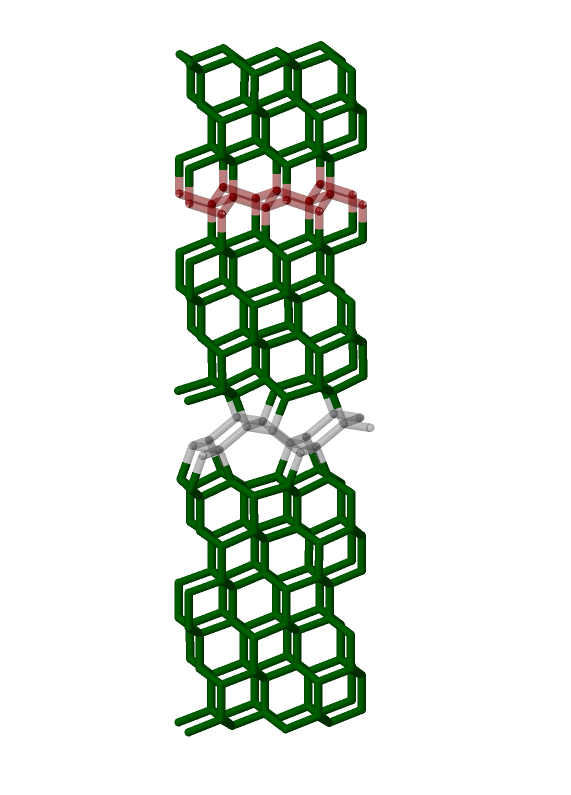}
\put(10,90){\fcolorbox{black}{white}{\Large a)}}
\end{overpic}
\begin{overpic}[width = 0.49\textwidth]{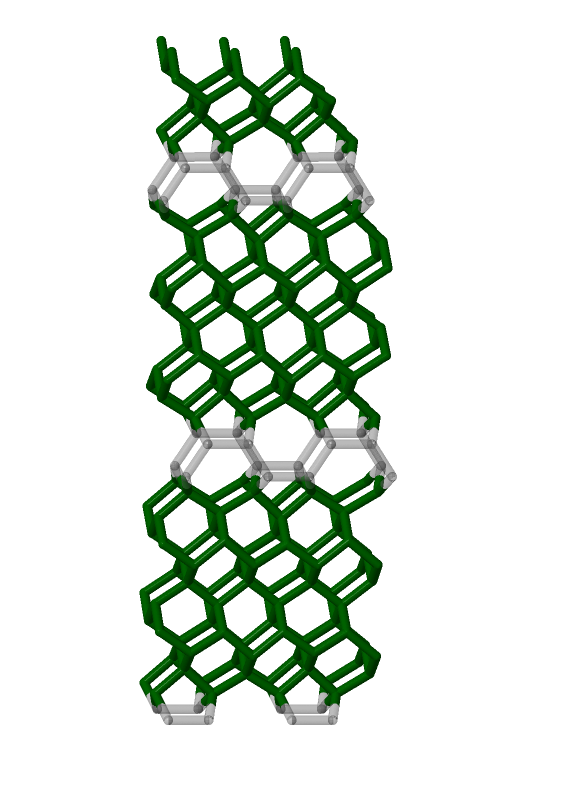}
\put(10,90){\fcolorbox{black}{white}{\Large b)}}
\end{overpic}
\label{f:2}
\end{figure*}

These two proposed earlier defect structures \cite{kondrin:drm20} are depicted in Fig.~\ref{f:2}. They can be regarded as two ``crumpled'' monoatomic layers, aligned along (111) and (100) direction (so they were named (111) and (100) defect structures respectively), separated by slabs of pristine crystal structures. They are related to  M-carbon \cite{oganov:jcp06,li:prl09} and S-carbon \cite{he:ssc12}. The last (100) structure of planar defect in diamond  was considered before \cite{goss:prb06}. Computer simulation of amorphous carbon and its subsequent deconvolution into ``basic'' crystal structures demonstrated, that combination of crumpled layers with 5-, 7- membered cycles (similar to defects  considered here) and pristine diamond/lonsdeleite diamond slabs  has the lowest energy \cite{deringer:cpc17}. In our consideration it is important to keep in mind, that the defect itself (crystal layers located in between two interface sheets shown in Fig.~\ref{f:2} by translucent color) may be shifted with respect to the surrounding crystal. On the other hand, pristine crystal slabs separated by the defect should not shift relative to each other, so their position remains the same as in initial pure crystal. In other words, thin planar defect (consisting of few atomic layers) can be created by finite transverse shift in the bulk of infinitely thick crystal by not perturbing the relative position of atomic layers in the pristine crystal. This requirement allows to create infinitesimal number of thin defects in the limit of low temperatures, when the defect concentration is low. Another important note is, that the two interface sheets are identical  in the case of (100) defect and different in case of (111) defect. This observation is related to different crystal symmetry of the corresponding defect structures -- it is monoclinic in case of (111) defect and orthorhombic in case of (100) one. 

As it was demonstrated in Ref.~\cite{kondrin:pssb22}, the concentration of interface sheets is described by the formula:

\begin{equation}
x=\left(\exp\left(\frac{\Delta E_1 + \Delta E_2}{k_B T}\right)+1\right)^{-1}
\label{eq:x}
\end{equation}

where $\Delta E_i$ is a formation energy of one of interface sheets per the number of atoms in the sheet. Note that concentration of interface sheets in the limit of infinitely high temperature converges to 0.5, that is, to evenly mixed layers  of pristine crystal  structure, separated by interface atomic sheets.

From concentration of the defects, we can calculate their contribution  to various thermodynamic properties of the defected crystal:
\begin{equation}
\Delta c_P=(\Delta E_1+\Delta E_2)\frac{dx}{dT}
\end{equation} 
and 
\begin{equation}
\Delta \alpha_V=\frac{\Delta V_1+\Delta V_2}{V} \frac{dx}{dT}
\label{eq:v}
\end{equation}

 For this purpose we have to know only the formation energy of the defect $\Delta E_i$ and its effect on the specific volume of the crystal $\Delta V_i$. We should stress one more time, that these planar defects are purely virtual and can be regarded as another way to treat explicit anharmonicity of the crystal. In this way, anharmonic contribution to the thermodynamic properties acquires obvious meaning as a change of crystal energy  and volume caused by the defects. Note, that calculation of these properties does  require nothing more except the ab-initio calculations at zero temperature.

It is interesting to note, that beforehand we don't know, which type of anharmonicity (bulk or transverse one) produces higher effect on thermodynamic properties of particular crystal. It seems that bulk input dominates in metals. See e.g. Ref.~\cite{grabowski:prb09}, where  it was demonstrated, that high-temperature specific heat and expansion coefficient in aluminum are mainly described by bulk anharmonicity. It seems, that in covalently bonded materials the situation is opposite, and the large part of anharmonicity present in them is due to transverse anharmonicity (like in germanium \cite{kondrin:pssb22}). More than half a century ago, there was a discussion about ``microscopic and macroscopic'' expansion of solids measured by different methods (by macroscopic dilatometry and microscopic X-Ray diffraction, see e.g. \cite{dutta:pssb62}). Involuntarily, we are about to provide another (thermodynamic) dimension for this discussion , as the difference between bulk and transverse anharmonicity. 

\section{Methods}
To calculate the formation energy of defect silicon structure we used  QuantumESPRESSO software package\cite{giannozzi:jopcm17}. For preparation of input files we heavily used \verb|cif2cell| program \cite{cif2cell}.
 
For the density functional calculation of formation energy, we employed the Perdew-Burke-Ernzerhof exchange correlation method with norm-conserving pseudopotentials with energy cutoff 70 Ry. The planar  defects were simulated in approximately 48-atom supercell. For integration over Brillouin zone, unshifted $4 \times 8 \times 1$ Monkhorst-Pack grid was used (with the number of nodes being roughly inversely proportional to the length in a corresponding direction). Crystal lattices and atom positions at fixed external pressure were fully optimized, until residual force on every atom did not exceed 0.001 Ry/bohr and additional stress -- 0.5 kbar. The estimated convergence error does not exceed 2 meV/atom. The formation energy of the defect was calculated by the energy difference between defected structure ($E$) and  defect-free silicon  lattice ($E_{host}$) of  similar dimensions:
\begin{equation} 
\Delta E = E-E_{host} n/n_{host}
\end{equation} 
where $n$ and $n_{host}$ are the number of atoms in defected and pristine silicon. It should be noted, that for planar defects, not the  formation energy defined this way, but rather its  normalized value (by the  layer interface area or the number of atoms, comprising the interface layers) has physical meaning. We will call it a specific formation energy.

\section{Results}
The specific formation energy $\Delta E_1+\Delta E_2$ is equal to 190 meV/atom (29 meV//\AA$^2$)) for (111) defect and 355 meV/atom (48 meV/\AA$^2$) for (100) defect. Interesting to note that despite this strong difference between respective energies per interface atom due to strong asymmetry of (111) defect the ``upper limit on the melting temperature'' (the energy barrier which  has to be overcame by shear displacement of atomic layers with respect to one another \cite{kondrin:drm20}) set by both types of defects is about the same. Since the one of the interface sheets of (111) defect consists of purely lonsdeleite type of interatomic bonds (see Fig.~\ref{f:2}), whose energy is close to that of pristine silicon. Consequently, almost all of additional energy in this defect is introduced by exactly  half of interface atoms. So, this upper limit (190 meV/atom $\approx$ 2190 K/atom) is only slightly larger than experimentally observed melting temperature (1690 K).

The additional volume, introduced by defects, requires slightly subtler calculations. The defects with concentration of interface sheets equal to $x=1/6$ that depicted in Fig.~\ref{f:2} have relative volume expansion  $\frac{\Delta V}{x V}$ (which has to be substituted in Eq.~\ref{eq:v}) equal to 0.03642 (for (111) defect) and 0.08439 (for (100) defect). However, in this calculation it was implicitly assumed, that all defects are aligned along the single crystallographic direction, and two types of  defects are formed independently of  each other. Therefore, we have to divide relative volume expansion by the number of equivalent crystallographic direction for respective defect type (4 for (111) defect and 3 for (100) defect). To take into account the difference between defects, we assume, that they independently occupy the crystal volume, so, the volume is equally distributed by the two types of defects ( corresponding ``occupation factor'' is equal to 0.5 for both defects). For simplicity, we also assume that change of volume is caused by the change of lengths in the direction, normal to interface sheets. This can be explained by that in the  low temperature limit (low defect concentrations) the area along the defect planes is determined by the area of rather thick layers of pristine silicon, which more or less retains geometry of pristine silicon crystal, and all changes in volume are restricted to the direction, normal to planar defects. This observation is corroborated by inspection of relative crystal lattice change in (111) defect with concentration 1/6. Taking  into account all these considerations, we can write down the final equation for linear thermal expansion along the single  (111) direction (the value 1100 in the formula $\approx 2196/2$):

\begin{equation}
\Delta l/l(111)= 0.5 \cdot 0.036 \cdot 1100/(2  T^2  \cosh(1100/T)^2)/4
\label{eq:111}
\end{equation} 

The similar equation with obvious corrections can be applied to linear expansion along (100) direction.

\begin{figure}
\includegraphics[width=\columnwidth]{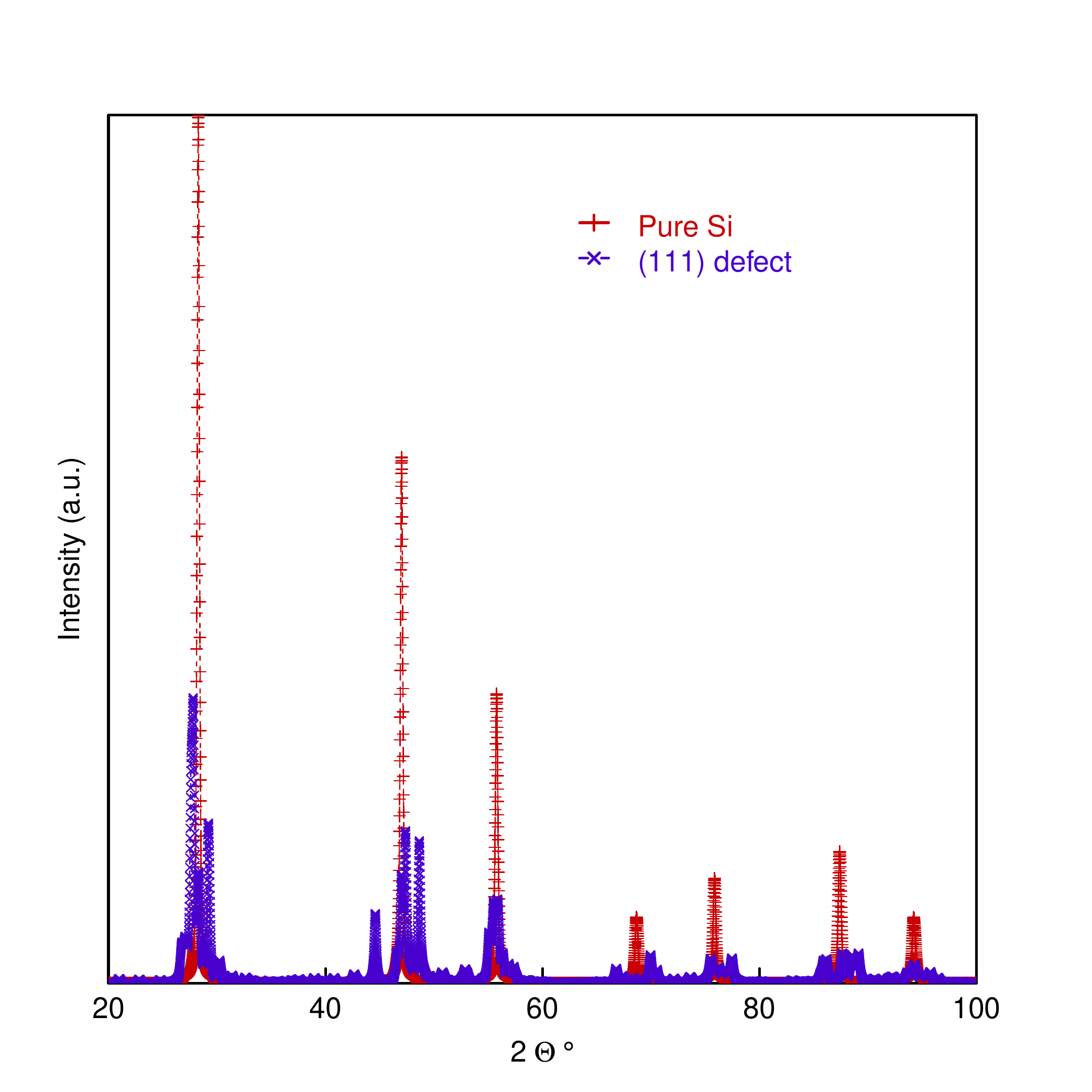}
\caption{Xray diffraction profiles for pristine silicon (red $+$) and silicon with (111) planar defects with concentration $x=1/6$ (blue $\times$). X-ray wavelength was assumed to be KCu$_\alpha$.}
\label{f:3}
\end{figure}

Eq.~\ref{eq:111} is depicted in Fig.~\ref{f:1} as thick green solid curve. For drawing this line we also include uniform bulk contribution from  the softening of interatomic bonds (shown as red dashed curve in Fig.~\ref{f:1}). We have compared it with experimental data of Maissel \cite{maissel:jap60} , where expansion coefficient along (111) direction of single crystal of silicon was measured by optical methods. We can see good coincidence of amplitude of predicted and measured expansion coefficient, and especially, correspondence of position of maximum on the expansion coefficient (at $\approx$ 900 K) observed for measured and calculated curves. Note also the difference between the values of linear expansion, measured by optical dilatometer and X-ray method. We argue, that X-ray method measures only uniform dilatation, caused by bulk anharmonicity although by the way it somewhat underestimates it. On the other hand, direct optical methods measure ``true'' dilatation, which also includes contribution from transverse anharmonic modes. The difference between the  two methods is evident if one draws X-Ray diffraction patterns from pristine silicon and silicon with (111) planar defects with concentration $x=1/6$ (see Fig.~\ref{f:3}). It can be easily concluded from inspection of his Figure, that accumulation of planar defects  leads to splitting of X-ray reflections   and eventually their broadening, rather than shift.  Note, that exact values of splitting and relative amplitudes of individual peaks are rather spurious issue  because they heavily depend on exact position of interface layers, so only the relative broadening of reflections which takes into account the related groups of peaks is meaningful in this case. Because of incoherent character of diffraction from planar defects randomly distributed along the bulk of the crystal, they can be compared to stacking faults and twins \cite{balogh:jap06} leading to broadening not the whole Bragg peak but rather of its ``pedestal''. Over this base the comparatively narrow coherent reflection is superimposed.  We should also remind, that significant broadening of optical phonon dispersion branches  measured by inelastic neutron scattering \cite{kim:prb15} as well as optically active Raman modes \cite{hart:prb70} were previously registered in experiments. So, we insist that it is the relative broadening of X-Ray peaks rather than change of  their position, that enables one to figure out the thermal expansion due to anharmonicity of transverse phonon modes. It should be taken into account , that concentration of interface layers along one direction used for simulation of X-ray diffraction pattern in Fig.~\ref{f:3} corresponds to very high temperature above the experimental melting temperature of silicon.

\begin{figure}
\includegraphics[width=\columnwidth]{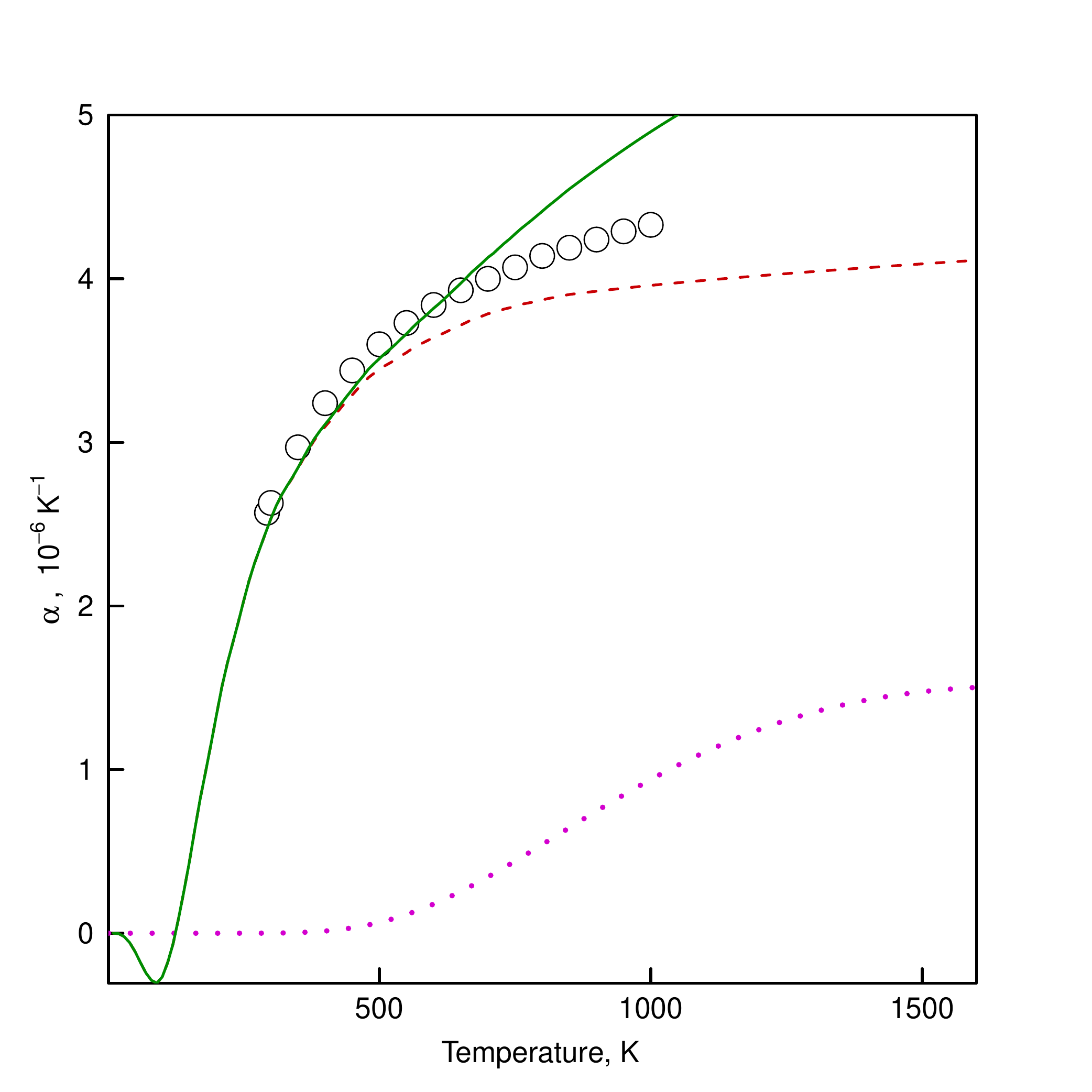}
\caption{Various theoretical and experimental data of linear expansion coefficient of silicon. Red dashed curve -- quasiharmonic approximation uniform expansion coefficient given by Eq:~\ref{eq:a1}. $\circ$ -- experimental data of Watanabe {\em et al.} \cite{watanabe:ijt04} measured along (100) direction by optical technique. Magenta dotted curve -- theoretical contribution from (100) planar defects. Green solid curve -- sum of two inputs to linear expansion coefficient from uniform dilatation and expansion caused by (100) planar defects. Note that the last curve is not a fit of experimental data because it contains no fitting parameters bat rather a comparison of theoretical and experimental results.}
\label{f:4}
\end{figure}

In similar way can be handled thermal expansion along (100) direction (see Fig.~\ref{f:4}). In this case for comparison we have chosen the most recent optical data of Watanabe et al. \cite{watanabe:ijt04}. It is clearly seen, that our approximation significantly exceeds experimental expansion coefficient, especially in the high temperature region. Still, for making the final decision about validity of our theory and its compliance to experiment, even more high temperature experimental data above 1000 K is needed. Up to now, to the best of our knowledge, no optical measurements of expansion coefficient of silicon to the temperatures up to the melting point exist. 

\begin{figure}
\includegraphics[width=\columnwidth]{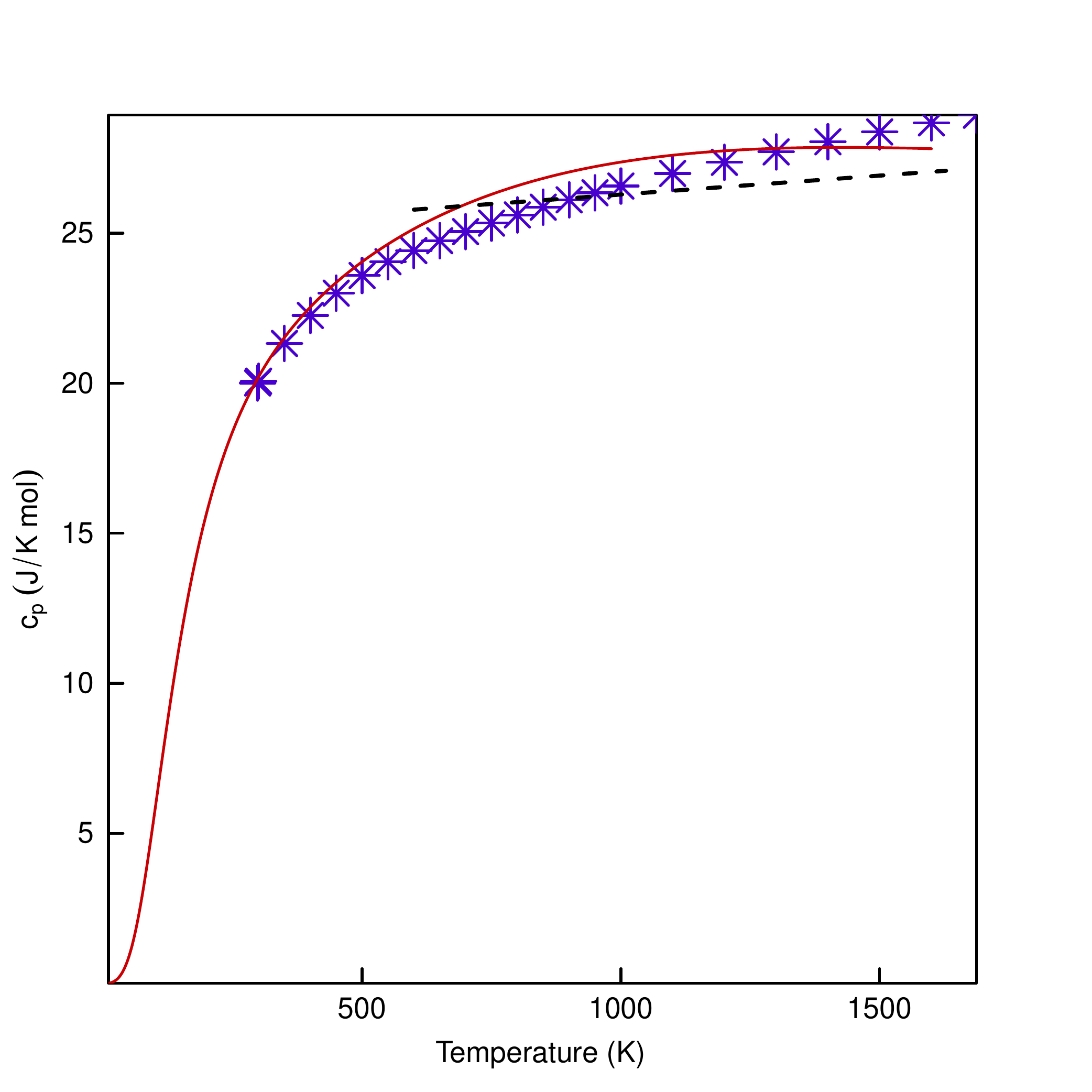}
\caption{Isobaric heat capacity of silicon. $\ast$ -- recommended experimental data of Desai \cite{desai:jpcrd86}. Red curve -- theoretical result including Debye contribution (with T$_D$=640 K and anharmonic input from two types of planar defects (111) and (100). Dashed black line -- calculation with Morse potential by Trivedi {\em et al.} \cite{trivedi:jpcssp77}.  Note that the last two curves are not fits of experimental data because they contain no fitting parameters bat rather a comparison of theoretical and experimental results.}
\label{f:5}
\end{figure}

Using relative concentration of (111) and (100) defects and their respective specific formation energy we can also evaluate isobaric specific heat (see Fig.~\ref{f:5}). Beside anharmonic contribution we added harmonic input described by Debye equation with T$_D$=640 K. For comparison we used the recommended experimental data of Desai \cite{desai:jpcrd86}. Experimental dependence is more or less reproduced by theoretical law with exception of high-temperature range where calculated data are below the experimental ones. The difference between high temperature data can be well explained by  presence of small concentration of even more high-energy planar defects namely (110) one \cite{kondrin:drm20} with specific formation energy 1.5 times higher, than that of (100) defect. We can also mention, that newer data of specific heat obtained by drop calorimeter methods \cite{yamaguchi:jtac02} produce substantially lower specific heat values. Earlier we have explained this underestimation of high temperature specific heat measured by drop calorimeters by the effect of planar defects quenching during cooling \cite{kondrin:pssb22}. The most reliable method of determination of specific heat would be ``one-shot measurement'' like pulse-heating one \cite{savva:pu20,savva:c16,kondratyev:prl19}. 

\section{Discussion}

We would like to briefly discuss the similarities and differences between our approach and s-TDEP method used earlier for determination of thermal expansion of silicon \cite{kim:pnas18}. Both methods are fully ab-initio zero temperature ones. The difference between them is that s-TDEP method used  stochastic sampling of 3$N$ coordinated (where $N$ is the number of atoms in calculation supercell) potential surface and subsequent fitting of obtained PES by third-order polynomial. We believe, that although this method by itself is sound,  there are some implicit assumptions in its fitting procedure. This procedure takes into account only two-phonon interactions and neglects higher order ones. Still, from this fitting and subsequent perturbative methods, it is very difficult to obtain anharmonic contribution to specific heat. Moreover, as stated in Ref.~\cite{cowley:rpp68}, the lowest order and highest order contributions to specific heat have large amplitude but different signs. Thus, the series are slow to converge.

As a side remark> it is interesting to recall rather old research of Trivedi {\em et al.}  \cite{trivedi:jpcssp77}. They calculate high temperature anharmonic contribution to specific heat of elements with diamond structure, taking into account only terms up to fourth order, but with artificially chosen interatomic potential (namely the Morse potential). The result of their calculation is shown in Fig.~\ref{f:5}. It is clear, that their calculations  heavily underestimate the trend, observed in experiments. We believe, that this discrepancy is due to the fact, that the Morse potential significantly underestimates anharmonicity present in real silicon.  

On the other hand, our approach is not perturbative and takes into account only the modes, producing metastable local minima at finite displacement with quite low energies. In other words, we assume, that the strongest input into anharmonic thermodynamic properties is due to ``topologically'' different modes (having different number of maxima/minima from harmonic potential). In some detail it was discussed in Ref.~\cite{kondrin:pssb22}. Despite not very strict formulation of the method, these local minima can be found as the energy lowering,  produced by bond reconstruction. Taking into account the great success recently achieved by various methods of {\em ab-initio} search for new crystal structures \cite{oganov:jcp06}, this approach (as we believe) can be easily computerized. Example of Ref.~\cite{deringer:cpc17}, where two of planar defects in diamonds similar to the one considered here was found by purely computer procedure, convinces us in eligibility of such an approach.

In some extent, this approach can be compared to the one taken earlier \cite{hart:prb70,menendez:prb84,debernardi:prl95}, where temperature dependence of Raman mode linewidth in silicon was interpreted in terms of anharmonic decay of $\Gamma$ optical phonon into two acoustical phonons with opposite wavevectors. It was shown earlier by measuring second order Raman under high temperatures \cite{klotz:prl97,brazhkin:jetpl00} and ultrasonic measurements of softening of phonon modes under high pressure \cite{goncharova:ftt83en}, that it is the transverse acoustic phonon mode at the edge of Brillouine zone in X point, which is most influenced by anharmonic effects. This anharmonicity is in some way transferred to optical phonons, which are prone to decay into this (transverse acoustical) channel. 

From more philosophical standing point, duality anharmonicity-defect can be regarded as an extension of de Broigle's duality wave-particle to phonons. As with fundamental particles, the particle-like  properties of phonons are realized, then there is an inter-phonon coupling that is anharmonicity, which in its turn leads to formation of crystal defects. It is like an electron, which acquires definite coordinate only when  interacting with detector material . Although, we can not provide direct conclusions from this duality, but it may be interesting  to dig into this analogy deeper. 

\section{Conclusions}  

Using analogy between phonon anharmonicity and crystal defects and the result of rather inexpensive zero temperature {\em ab-initio} calculations, we are able to describe high-temperature thermal expansion coefficient and isobaric specific heat of silicon. We have demonstrated, that anharmonicity of only two transverse phonon modes (propagating along (111) and (100) directions)  contributes to these thermodynamic properties at high-temperatures. This calculations accounts for more than 10\% deviation of experimentally measured thermodynamic properties in the vicinity of melting point of silicon from the values obtained by quasi-harmonic approximation. 

\section*{Acknowledgements}
This work is supported by the Russian Science Foundation (grant No. 19-12-00111).

\begin{thebibliography}{40}%
\makeatletter
\providecommand \@ifxundefined [1]{%
 \@ifx{#1\undefined}
}%
\providecommand \@ifnum [1]{%
 \ifnum #1\expandafter \@firstoftwo
 \else \expandafter \@secondoftwo
 \fi
}%
\providecommand \@ifx [1]{%
 \ifx #1\expandafter \@firstoftwo
 \else \expandafter \@secondoftwo
 \fi
}%
\providecommand \natexlab [1]{#1}%
\providecommand \enquote  [1]{``#1''}%
\providecommand \bibnamefont  [1]{#1}%
\providecommand \bibfnamefont [1]{#1}%
\providecommand \citenamefont [1]{#1}%
\providecommand \href@noop [0]{\@secondoftwo}%
\providecommand \href [0]{\begingroup \@sanitize@url \@href}%
\providecommand \@href[1]{\@@startlink{#1}\@@href}%
\providecommand \@@href[1]{\endgroup#1\@@endlink}%
\providecommand \@sanitize@url [0]{\catcode `\\12\catcode `\$12\catcode
  `\&12\catcode `\#12\catcode `\^12\catcode `\_12\catcode `\%12\relax}%
\providecommand \@@startlink[1]{}%
\providecommand \@@endlink[0]{}%
\providecommand \url  [0]{\begingroup\@sanitize@url \@url }%
\providecommand \@url [1]{\endgroup\@href {#1}{\urlprefix }}%
\providecommand \urlprefix  [0]{URL }%
\providecommand \Eprint [0]{\href }%
\providecommand \doibase [0]{http://dx.doi.org/}%
\providecommand \selectlanguage [0]{\@gobble}%
\providecommand \bibinfo  [0]{\@secondoftwo}%
\providecommand \bibfield  [0]{\@secondoftwo}%
\providecommand \translation [1]{[#1]}%
\providecommand \BibitemOpen [0]{}%
\providecommand \bibitemStop [0]{}%
\providecommand \bibitemNoStop [0]{.\EOS\space}%
\providecommand \EOS [0]{\spacefactor3000\relax}%
\providecommand \BibitemShut  [1]{\csname bibitem#1\endcsname}%
\let\auto@bib@innerbib\@empty
\bibitem [{\citenamefont {Landau}\ \emph {et~al.}(1980)\citenamefont {Landau},
  \citenamefont {Pitaevskii},\ and\ \citenamefont {Lifshitz}}]{landafshitz:v}%
  \BibitemOpen
  \bibfield  {author} {\bibinfo {author} {\bibfnamefont {L.}~\bibnamefont
  {Landau}}, \bibinfo {author} {\bibfnamefont {L.}~\bibnamefont {Pitaevskii}},
  \ and\ \bibinfo {author} {\bibfnamefont {E.}~\bibnamefont {Lifshitz}},\
  }\href@noop {} {\emph {\bibinfo {title} {Statistical Physics}}},\ Course of
  theoretical physics\ (\bibinfo  {publisher} {Pergamon Press, Oxford},\
  \bibinfo {year} {1980})\BibitemShut {NoStop}%
\bibitem [{\citenamefont {Swenson}(1983)}]{swenson:jpcrd83}%
  \BibitemOpen
  \bibfield  {author} {\bibinfo {author} {\bibfnamefont {C.~A.}\ \bibnamefont
  {Swenson}},\ }\href {\doibase 10.1063/1.555681} {\bibfield  {journal}
  {\bibinfo  {journal} {Journal of Physical and Chemical Reference Data}\
  }\textbf {\bibinfo {volume} {12}},\ \bibinfo {pages} {179} (\bibinfo {year}
  {1983})}\BibitemShut {NoStop}%
\bibitem [{\citenamefont {Born}\ and\ \citenamefont {Brody}(1921)}]{born:zp21}%
  \BibitemOpen
  \bibfield  {author} {\bibinfo {author} {\bibfnamefont {M.}~\bibnamefont
  {Born}}\ and\ \bibinfo {author} {\bibfnamefont {E.}~\bibnamefont {Brody}},\
  }\href {\doibase 10.1007/BF01327972} {\bibfield  {journal} {\bibinfo
  {journal} {Zeitschrift f\"{u}r Physik}\ }\textbf {\bibinfo {volume} {6}},\
  \bibinfo {pages} {132} (\bibinfo {year} {1921})}\BibitemShut {NoStop}%
\bibitem [{\citenamefont {Wallace}(1965)}]{wallace:pr65}%
  \BibitemOpen
  \bibfield  {author} {\bibinfo {author} {\bibfnamefont {D.~C.}\ \bibnamefont
  {Wallace}},\ }\href {\doibase 10.1103/PhysRev.139.A877} {\bibfield  {journal}
  {\bibinfo  {journal} {Phys. Rev.}\ }\textbf {\bibinfo {volume} {139}},\
  \bibinfo {pages} {A877} (\bibinfo {year} {1965})}\BibitemShut {NoStop}%
\bibitem [{\citenamefont {Cowley}(1968)}]{cowley:rpp68}%
  \BibitemOpen
  \bibfield  {author} {\bibinfo {author} {\bibfnamefont {R.~A.}\ \bibnamefont
  {Cowley}},\ }\href {\doibase 10.1088/0034-4885/31/1/303} {\bibfield
  {journal} {\bibinfo  {journal} {Reports on Progress in Physics}\ }\textbf
  {\bibinfo {volume} {31}},\ \bibinfo {pages} {123} (\bibinfo {year}
  {1968})}\BibitemShut {NoStop}%
\bibitem [{\citenamefont {Gerlich}\ \emph {et~al.}(1965)\citenamefont
  {Gerlich}, \citenamefont {Abeles},\ and\ \citenamefont
  {Miller}}]{gerlich:jap65}%
  \BibitemOpen
  \bibfield  {author} {\bibinfo {author} {\bibfnamefont {D.}~\bibnamefont
  {Gerlich}}, \bibinfo {author} {\bibfnamefont {B.}~\bibnamefont {Abeles}}, \
  and\ \bibinfo {author} {\bibfnamefont {R.~E.}\ \bibnamefont {Miller}},\
  }\href {\doibase 10.1063/1.1713926} {\bibfield  {journal} {\bibinfo
  {journal} {Journal of Applied Physics}\ }\textbf {\bibinfo {volume} {36}},\
  \bibinfo {pages} {76} (\bibinfo {year} {1965})}\BibitemShut {NoStop}%
\bibitem [{\citenamefont {Desai}(1986)}]{desai:jpcrd86}%
  \BibitemOpen
  \bibfield  {author} {\bibinfo {author} {\bibfnamefont {P.~D.}\ \bibnamefont
  {Desai}},\ }\href {\doibase 10.1063/1.555761} {\bibfield  {journal} {\bibinfo
   {journal} {Journal of Physical and Chemical Reference Data}\ }\textbf
  {\bibinfo {volume} {15}},\ \bibinfo {pages} {967} (\bibinfo {year}
  {1986})}\BibitemShut {NoStop}%
\bibitem [{\citenamefont {Yamaguchi}\ and\ \citenamefont
  {Itagaki}(2002)}]{yamaguchi:jtac02}%
  \BibitemOpen
  \bibfield  {author} {\bibinfo {author} {\bibfnamefont {K.}~\bibnamefont
  {Yamaguchi}}\ and\ \bibinfo {author} {\bibfnamefont {K.}~\bibnamefont
  {Itagaki}},\ }\href {\doibase 10.1023/A:1020609517891} {\bibfield  {journal}
  {\bibinfo  {journal} {Journal of Thermal Analysis and Calorimetry}\ }\textbf
  {\bibinfo {volume} {69}},\ \bibinfo {pages} {1059} (\bibinfo {year}
  {2002})}\BibitemShut {NoStop}%
\bibitem [{\citenamefont {Maissel}(1960)}]{maissel:jap60}%
  \BibitemOpen
  \bibfield  {author} {\bibinfo {author} {\bibfnamefont {L.}~\bibnamefont
  {Maissel}},\ }\href {\doibase 10.1063/1.1735401} {\bibfield  {journal}
  {\bibinfo  {journal} {Journal of Applied Physics}\ }\textbf {\bibinfo
  {volume} {31}},\ \bibinfo {pages} {211} (\bibinfo {year} {1960})}\BibitemShut
  {NoStop}%
\bibitem [{\citenamefont {Watanabe}\ \emph {et~al.}(2004)\citenamefont
  {Watanabe}, \citenamefont {Yamada},\ and\ \citenamefont
  {Okaji}}]{watanabe:ijt04}%
  \BibitemOpen
  \bibfield  {author} {\bibinfo {author} {\bibfnamefont {H.}~\bibnamefont
  {Watanabe}}, \bibinfo {author} {\bibfnamefont {N.}~\bibnamefont {Yamada}}, \
  and\ \bibinfo {author} {\bibfnamefont {M.}~\bibnamefont {Okaji}},\ }\href
  {\doibase 10.1023/B:IJOT.0000022336.83719.43} {\bibfield  {journal} {\bibinfo
   {journal} {International Journal of Thermophysics}\ }\textbf {\bibinfo
  {volume} {25}},\ \bibinfo {pages} {221} (\bibinfo {year} {2004})}\BibitemShut
  {NoStop}%
\bibitem [{\citenamefont {Dutta}(1962)}]{dutta:pssb62}%
  \BibitemOpen
  \bibfield  {author} {\bibinfo {author} {\bibfnamefont {B.~N.}\ \bibnamefont
  {Dutta}},\ }\href {\doibase https://doi.org/10.1002/pssb.19620020803}
  {\bibfield  {journal} {\bibinfo  {journal} {physica status solidi (b)}\
  }\textbf {\bibinfo {volume} {2}},\ \bibinfo {pages} {984} (\bibinfo {year}
  {1962})}\BibitemShut {NoStop}%
\bibitem [{\citenamefont {Okada}\ and\ \citenamefont
  {Tokumaru}(1984)}]{okada:jap84}%
  \BibitemOpen
  \bibfield  {author} {\bibinfo {author} {\bibfnamefont {Y.}~\bibnamefont
  {Okada}}\ and\ \bibinfo {author} {\bibfnamefont {Y.}~\bibnamefont
  {Tokumaru}},\ }\href {\doibase 10.1063/1.333965} {\bibfield  {journal}
  {\bibinfo  {journal} {Journal of Applied Physics}\ }\textbf {\bibinfo
  {volume} {56}},\ \bibinfo {pages} {314} (\bibinfo {year} {1984})}\BibitemShut
  {NoStop}%
\bibitem [{\citenamefont {Roberts}(1981)}]{roberts:jpdap81}%
  \BibitemOpen
  \bibfield  {author} {\bibinfo {author} {\bibfnamefont {R.~B.}\ \bibnamefont
  {Roberts}},\ }\href {\doibase 10.1088/0022-3727/14/10/003} {\bibfield
  {journal} {\bibinfo  {journal} {Journal of Physics D: Applied Physics}\
  }\textbf {\bibinfo {volume} {14}},\ \bibinfo {pages} {L163} (\bibinfo {year}
  {1981})}\BibitemShut {NoStop}%
\bibitem [{\citenamefont {Grabowski}\ \emph {et~al.}(2009)\citenamefont
  {Grabowski}, \citenamefont {Ismer}, \citenamefont {Hickel},\ and\
  \citenamefont {Neugebauer}}]{grabowski:prb09}%
  \BibitemOpen
  \bibfield  {author} {\bibinfo {author} {\bibfnamefont {B.}~\bibnamefont
  {Grabowski}}, \bibinfo {author} {\bibfnamefont {L.}~\bibnamefont {Ismer}},
  \bibinfo {author} {\bibfnamefont {T.}~\bibnamefont {Hickel}}, \ and\ \bibinfo
  {author} {\bibfnamefont {J.}~\bibnamefont {Neugebauer}},\ }\href {\doibase
  10.1103/PhysRevB.79.134106} {\bibfield  {journal} {\bibinfo  {journal} {Phys.
  Rev. B}\ }\textbf {\bibinfo {volume} {79}},\ \bibinfo {pages} {134106}
  (\bibinfo {year} {2009})}\BibitemShut {NoStop}%
\bibitem [{\citenamefont {Kim}\ \emph {et~al.}(2018)\citenamefont {Kim},
  \citenamefont {Hellman}, \citenamefont {Herriman}, \citenamefont {Smith},
  \citenamefont {Lin}, \citenamefont {Shulumba}, \citenamefont {Niedziela},
  \citenamefont {Li}, \citenamefont {Abernathy},\ and\ \citenamefont
  {Fultz}}]{kim:pnas18}%
  \BibitemOpen
  \bibfield  {author} {\bibinfo {author} {\bibfnamefont {D.~S.}\ \bibnamefont
  {Kim}}, \bibinfo {author} {\bibfnamefont {O.}~\bibnamefont {Hellman}},
  \bibinfo {author} {\bibfnamefont {J.}~\bibnamefont {Herriman}}, \bibinfo
  {author} {\bibfnamefont {H.~L.}\ \bibnamefont {Smith}}, \bibinfo {author}
  {\bibfnamefont {J.~Y.~Y.}\ \bibnamefont {Lin}}, \bibinfo {author}
  {\bibfnamefont {N.}~\bibnamefont {Shulumba}}, \bibinfo {author}
  {\bibfnamefont {J.~L.}\ \bibnamefont {Niedziela}}, \bibinfo {author}
  {\bibfnamefont {C.~W.}\ \bibnamefont {Li}}, \bibinfo {author} {\bibfnamefont
  {D.~L.}\ \bibnamefont {Abernathy}}, \ and\ \bibinfo {author} {\bibfnamefont
  {B.}~\bibnamefont {Fultz}},\ }\href {\doibase 10.1073/pnas.1707745115}
  {\bibfield  {journal} {\bibinfo  {journal} {Proceedings of the National
  Academy of Sciences}\ }\textbf {\bibinfo {volume} {115}},\ \bibinfo {pages}
  {1992} (\bibinfo {year} {2018})}\BibitemShut {NoStop}%
\bibitem [{\citenamefont {Kondrin}\ \emph {et~al.}(2020)\citenamefont
  {Kondrin}, \citenamefont {Lebed},\ and\ \citenamefont
  {Brazhkin}}]{kondrin:drm20}%
  \BibitemOpen
  \bibfield  {author} {\bibinfo {author} {\bibfnamefont {M.}~\bibnamefont
  {Kondrin}}, \bibinfo {author} {\bibfnamefont {Y.}~\bibnamefont {Lebed}}, \
  and\ \bibinfo {author} {\bibfnamefont {V.}~\bibnamefont {Brazhkin}},\ }\href
  {\doibase https://doi.org/10.1016/j.diamond.2020.108114} {\bibfield
  {journal} {\bibinfo  {journal} {Diamond and Related Materials}\ }\textbf
  {\bibinfo {volume} {110}},\ \bibinfo {pages} {108114} (\bibinfo {year}
  {2020})}\BibitemShut {NoStop}%
\bibitem [{\citenamefont {Kondrin}\ \emph {et~al.}(2021)\citenamefont
  {Kondrin}, \citenamefont {Lebed},\ and\ \citenamefont
  {Brazhkin}}]{kondrin:prl21}%
  \BibitemOpen
  \bibfield  {author} {\bibinfo {author} {\bibfnamefont {M.~V.}\ \bibnamefont
  {Kondrin}}, \bibinfo {author} {\bibfnamefont {Y.~B.}\ \bibnamefont {Lebed}},
  \ and\ \bibinfo {author} {\bibfnamefont {V.~V.}\ \bibnamefont {Brazhkin}},\
  }\href {\doibase 10.1103/PhysRevLett.126.165501} {\bibfield  {journal}
  {\bibinfo  {journal} {Phys. Rev. Lett.}\ }\textbf {\bibinfo {volume} {126}},\
  \bibinfo {pages} {165501} (\bibinfo {year} {2021})}\BibitemShut {NoStop}%
\bibitem [{\citenamefont {Kondrin}\ \emph {et~al.}(2022)\citenamefont
  {Kondrin}, \citenamefont {Lebed},\ and\ \citenamefont
  {Brazhkin}}]{kondrin:pssb22}%
  \BibitemOpen
  \bibfield  {author} {\bibinfo {author} {\bibfnamefont {M.}~\bibnamefont
  {Kondrin}}, \bibinfo {author} {\bibfnamefont {Y.}~\bibnamefont {Lebed}}, \
  and\ \bibinfo {author} {\bibfnamefont {V.}~\bibnamefont {Brazhkin}},\ }\href
  {\doibase 10.1002/pssb.202100463} {\bibfield  {journal} {\bibinfo  {journal}
  {physica status solidi (b)}\ }\textbf {\bibinfo {volume} {259}},\ \bibinfo
  {pages} {2100463} (\bibinfo {year} {2022})}\BibitemShut {NoStop}%
\bibitem [{\citenamefont {Savvatimskii}\ and\ \citenamefont
  {Onufriev}(2020)}]{savva:pu20}%
  \BibitemOpen
  \bibfield  {author} {\bibinfo {author} {\bibfnamefont {A.~I.}\ \bibnamefont
  {Savvatimskii}}\ and\ \bibinfo {author} {\bibfnamefont {S.~V.}\ \bibnamefont
  {Onufriev}},\ }\href {\doibase 10.3367/UFNe.2019.10.038665} {\bibfield
  {journal} {\bibinfo  {journal} {Phys. Usp.}\ }\textbf {\bibinfo {volume}
  {63}},\ \bibinfo {pages} {1015} (\bibinfo {year} {2020})}\BibitemShut
  {NoStop}%
\bibitem [{\citenamefont {Savvatimskiy}\ \emph {et~al.}(2016)\citenamefont
  {Savvatimskiy}, \citenamefont {Onufriev},\ and\ \citenamefont
  {Kondratyev}}]{savva:c16}%
  \BibitemOpen
  \bibfield  {author} {\bibinfo {author} {\bibfnamefont {A.}~\bibnamefont
  {Savvatimskiy}}, \bibinfo {author} {\bibfnamefont {S.}~\bibnamefont
  {Onufriev}}, \ and\ \bibinfo {author} {\bibfnamefont {A.}~\bibnamefont
  {Kondratyev}},\ }\href {\doibase
  https://doi.org/10.1016/j.carbon.2015.11.044} {\bibfield  {journal} {\bibinfo
   {journal} {Carbon}\ }\textbf {\bibinfo {volume} {98}},\ \bibinfo {pages}
  {534 } (\bibinfo {year} {2016})}\BibitemShut {NoStop}%
\bibitem [{\citenamefont {Kondratyev}\ and\ \citenamefont
  {Rakhel}(2019)}]{kondratyev:prl19}%
  \BibitemOpen
  \bibfield  {author} {\bibinfo {author} {\bibfnamefont {A.~M.}\ \bibnamefont
  {Kondratyev}}\ and\ \bibinfo {author} {\bibfnamefont {A.~D.}\ \bibnamefont
  {Rakhel}},\ }\href {\doibase 10.1103/PhysRevLett.122.175702} {\bibfield
  {journal} {\bibinfo  {journal} {Phys. Rev. Lett.}\ }\textbf {\bibinfo
  {volume} {122}},\ \bibinfo {pages} {175702} (\bibinfo {year}
  {2019})}\BibitemShut {NoStop}%
\bibitem [{\citenamefont {Vanhellemont}\ \emph {et~al.}(2014)\citenamefont
  {Vanhellemont}, \citenamefont {Swarnakar},\ and\ \citenamefont {der
  Biest}}]{vanhellemont:ecst14}%
  \BibitemOpen
  \bibfield  {author} {\bibinfo {author} {\bibfnamefont {J.}~\bibnamefont
  {Vanhellemont}}, \bibinfo {author} {\bibfnamefont {A.~K.}\ \bibnamefont
  {Swarnakar}}, \ and\ \bibinfo {author} {\bibfnamefont {O.~V.}\ \bibnamefont
  {der Biest}},\ }\href {\doibase 10.1149/06411.0283ecst} {\bibfield  {journal}
  {\bibinfo  {journal} {{ECS} Transactions}\ }\textbf {\bibinfo {volume}
  {64}},\ \bibinfo {pages} {283} (\bibinfo {year} {2014})}\BibitemShut
  {NoStop}%
\bibitem [{\citenamefont {Goncharova}\ \emph {et~al.}(1983)\citenamefont
  {Goncharova}, \citenamefont {Chernysheva},\ and\ \citenamefont
  {Voronov}}]{goncharova:ftt83en}%
  \BibitemOpen
  \bibfield  {author} {\bibinfo {author} {\bibfnamefont {V.~A.}\ \bibnamefont
  {Goncharova}}, \bibinfo {author} {\bibfnamefont {E.~V.}\ \bibnamefont
  {Chernysheva}}, \ and\ \bibinfo {author} {\bibfnamefont {F.~F.}\ \bibnamefont
  {Voronov}},\ }\href
  {http://www.mathnet.ru/php/archive.phtml?wshow=paper&jrnid=ftt&paperid=4249}
  {\bibfield  {journal} {\bibinfo  {journal} {Fizika Tverdogo Tela}\ }\textbf
  {\bibinfo {volume} {25}},\ \bibinfo {pages} {3680} (\bibinfo {year}
  {1983})},\ \bibinfo {note} {(in Russian)}\BibitemShut {NoStop}%
\bibitem [{\citenamefont {Kim}\ \emph {et~al.}(2015)\citenamefont {Kim},
  \citenamefont {Smith}, \citenamefont {Niedziela}, \citenamefont {Li},
  \citenamefont {Abernathy},\ and\ \citenamefont {Fultz}}]{kim:prb15}%
  \BibitemOpen
  \bibfield  {author} {\bibinfo {author} {\bibfnamefont {D.~S.}\ \bibnamefont
  {Kim}}, \bibinfo {author} {\bibfnamefont {H.~L.}\ \bibnamefont {Smith}},
  \bibinfo {author} {\bibfnamefont {J.~L.}\ \bibnamefont {Niedziela}}, \bibinfo
  {author} {\bibfnamefont {C.~W.}\ \bibnamefont {Li}}, \bibinfo {author}
  {\bibfnamefont {D.~L.}\ \bibnamefont {Abernathy}}, \ and\ \bibinfo {author}
  {\bibfnamefont {B.}~\bibnamefont {Fultz}},\ }\href {\doibase
  10.1103/PhysRevB.91.014307} {\bibfield  {journal} {\bibinfo  {journal} {Phys.
  Rev. B}\ }\textbf {\bibinfo {volume} {91}},\ \bibinfo {pages} {014307}
  (\bibinfo {year} {2015})}\BibitemShut {NoStop}%
\bibitem [{\citenamefont {Wei}\ \emph {et~al.}(1994)\citenamefont {Wei},
  \citenamefont {Li},\ and\ \citenamefont {Chou}}]{wei:prb94a}%
  \BibitemOpen
  \bibfield  {author} {\bibinfo {author} {\bibfnamefont {S.}~\bibnamefont
  {Wei}}, \bibinfo {author} {\bibfnamefont {C.}~\bibnamefont {Li}}, \ and\
  \bibinfo {author} {\bibfnamefont {M.~Y.}\ \bibnamefont {Chou}},\ }\href
  {\doibase 10.1103/PhysRevB.50.14587} {\bibfield  {journal} {\bibinfo
  {journal} {Phys. Rev. B}\ }\textbf {\bibinfo {volume} {50}},\ \bibinfo
  {pages} {14587} (\bibinfo {year} {1994})}\BibitemShut {NoStop}%
\bibitem [{\citenamefont {Wang}\ \emph {et~al.}(2015)\citenamefont {Wang},
  \citenamefont {Gu}, \citenamefont {Kuang},\ and\ \citenamefont
  {Xiang}}]{wang:zfna15}%
  \BibitemOpen
  \bibfield  {author} {\bibinfo {author} {\bibfnamefont {C.}~\bibnamefont
  {Wang}}, \bibinfo {author} {\bibfnamefont {J.}~\bibnamefont {Gu}}, \bibinfo
  {author} {\bibfnamefont {X.}~\bibnamefont {Kuang}}, \ and\ \bibinfo {author}
  {\bibfnamefont {S.}~\bibnamefont {Xiang}},\ }\href {\doibase
  10.1515/zna-2015-0027} {\bibfield  {journal} {\bibinfo  {journal}
  {Zeitschrift f\"{u}r Naturforschung A}\ }\textbf {\bibinfo {volume} {70}}
  (\bibinfo {year} {2015}),\ 10.1515/zna-2015-0027},\ \bibinfo {note}
  {\url{https://dx.doi.org/10.1515/zna-2015-0027}}\BibitemShut {NoStop}%
\bibitem [{\citenamefont {Oganov}\ and\ \citenamefont
  {Glass}(2006)}]{oganov:jcp06}%
  \BibitemOpen
  \bibfield  {author} {\bibinfo {author} {\bibfnamefont {A.~R.}\ \bibnamefont
  {Oganov}}\ and\ \bibinfo {author} {\bibfnamefont {C.~W.}\ \bibnamefont
  {Glass}},\ }\href {\doibase 10.1063/1.2210932} {\bibfield  {journal}
  {\bibinfo  {journal} {The Journal of Chemical Physics}\ }\textbf {\bibinfo
  {volume} {124}},\ \bibinfo {pages} {244704} (\bibinfo {year}
  {2006})}\BibitemShut {NoStop}%
\bibitem [{\citenamefont {Li}\ \emph {et~al.}(2009)\citenamefont {Li},
  \citenamefont {Ma}, \citenamefont {Oganov}, \citenamefont {Wang},
  \citenamefont {Wang}, \citenamefont {Xu}, \citenamefont {Cui}, \citenamefont
  {Mao},\ and\ \citenamefont {Zou}}]{li:prl09}%
  \BibitemOpen
  \bibfield  {author} {\bibinfo {author} {\bibfnamefont {Q.}~\bibnamefont
  {Li}}, \bibinfo {author} {\bibfnamefont {Y.}~\bibnamefont {Ma}}, \bibinfo
  {author} {\bibfnamefont {A.~R.}\ \bibnamefont {Oganov}}, \bibinfo {author}
  {\bibfnamefont {H.}~\bibnamefont {Wang}}, \bibinfo {author} {\bibfnamefont
  {H.}~\bibnamefont {Wang}}, \bibinfo {author} {\bibfnamefont {Y.}~\bibnamefont
  {Xu}}, \bibinfo {author} {\bibfnamefont {T.}~\bibnamefont {Cui}}, \bibinfo
  {author} {\bibfnamefont {H.-K.}\ \bibnamefont {Mao}}, \ and\ \bibinfo
  {author} {\bibfnamefont {G.}~\bibnamefont {Zou}},\ }\href {\doibase
  10.1103/PhysRevLett.102.175506} {\bibfield  {journal} {\bibinfo  {journal}
  {Phys. Rev. Lett.}\ }\textbf {\bibinfo {volume} {102}},\ \bibinfo {pages}
  {175506} (\bibinfo {year} {2009})}\BibitemShut {NoStop}%
\bibitem [{\citenamefont {He}\ \emph {et~al.}(2012)\citenamefont {He},
  \citenamefont {Sun}, \citenamefont {Zhang}, \citenamefont {Peng},
  \citenamefont {Zhang},\ and\ \citenamefont {Zhong}}]{he:ssc12}%
  \BibitemOpen
  \bibfield  {author} {\bibinfo {author} {\bibfnamefont {C.}~\bibnamefont
  {He}}, \bibinfo {author} {\bibfnamefont {L.}~\bibnamefont {Sun}}, \bibinfo
  {author} {\bibfnamefont {C.}~\bibnamefont {Zhang}}, \bibinfo {author}
  {\bibfnamefont {X.}~\bibnamefont {Peng}}, \bibinfo {author} {\bibfnamefont
  {K.}~\bibnamefont {Zhang}}, \ and\ \bibinfo {author} {\bibfnamefont
  {J.}~\bibnamefont {Zhong}},\ }\href {\doibase
  https://doi.org/10.1016/j.ssc.2012.05.022} {\bibfield  {journal} {\bibinfo
  {journal} {Solid State Communications}\ }\textbf {\bibinfo {volume} {152}},\
  \bibinfo {pages} {1560 } (\bibinfo {year} {2012})}\BibitemShut {NoStop}%
\bibitem [{\citenamefont {Goss}\ \emph {et~al.}(2006)\citenamefont {Goss},
  \citenamefont {Briddon}, \citenamefont {Jones},\ and\ \citenamefont
  {Heggie}}]{goss:prb06}%
  \BibitemOpen
  \bibfield  {author} {\bibinfo {author} {\bibfnamefont {J.~P.}\ \bibnamefont
  {Goss}}, \bibinfo {author} {\bibfnamefont {P.~R.}\ \bibnamefont {Briddon}},
  \bibinfo {author} {\bibfnamefont {R.}~\bibnamefont {Jones}}, \ and\ \bibinfo
  {author} {\bibfnamefont {M.~I.}\ \bibnamefont {Heggie}},\ }\href {\doibase
  10.1103/PhysRevB.73.115204} {\bibfield  {journal} {\bibinfo  {journal} {Phys.
  Rev. B}\ }\textbf {\bibinfo {volume} {73}},\ \bibinfo {pages} {115204}
  (\bibinfo {year} {2006})}\BibitemShut {NoStop}%
\bibitem [{\citenamefont {Deringer}\ \emph {et~al.}(2017)\citenamefont
  {Deringer}, \citenamefont {Cs\'{a}nyi},\ and\ \citenamefont
  {Proserpio}}]{deringer:cpc17}%
  \BibitemOpen
  \bibfield  {author} {\bibinfo {author} {\bibfnamefont {V.~L.}\ \bibnamefont
  {Deringer}}, \bibinfo {author} {\bibfnamefont {G.}~\bibnamefont
  {Cs\'{a}nyi}}, \ and\ \bibinfo {author} {\bibfnamefont {D.~M.}\ \bibnamefont
  {Proserpio}},\ }\href {\doibase 10.1002/cphc.201700151} {\bibfield  {journal}
  {\bibinfo  {journal} {ChemPhysChem}\ }\textbf {\bibinfo {volume} {18}},\
  \bibinfo {pages} {873} (\bibinfo {year} {2017})}\BibitemShut {NoStop}%
\bibitem [{\citenamefont {Giannozzi}\ \emph {et~al.}(2017)\citenamefont
  {Giannozzi}, \citenamefont {Andreussi}, \citenamefont {Brumme}, \citenamefont
  {Bunau}, \citenamefont {Nardelli}, \citenamefont {Calandra}, \citenamefont
  {Car}, \citenamefont {Cavazzoni}, \citenamefont {Ceresoli}, \citenamefont
  {Cococcioni},\ and\ \citenamefont {{\em et. al.}}}]{giannozzi:jopcm17}%
  \BibitemOpen
  \bibfield  {author} {\bibinfo {author} {\bibfnamefont {P.}~\bibnamefont
  {Giannozzi}}, \bibinfo {author} {\bibfnamefont {O.}~\bibnamefont
  {Andreussi}}, \bibinfo {author} {\bibfnamefont {T.}~\bibnamefont {Brumme}},
  \bibinfo {author} {\bibfnamefont {O.}~\bibnamefont {Bunau}}, \bibinfo
  {author} {\bibfnamefont {M.~B.}\ \bibnamefont {Nardelli}}, \bibinfo {author}
  {\bibfnamefont {M.}~\bibnamefont {Calandra}}, \bibinfo {author}
  {\bibfnamefont {R.}~\bibnamefont {Car}}, \bibinfo {author} {\bibfnamefont
  {C.}~\bibnamefont {Cavazzoni}}, \bibinfo {author} {\bibfnamefont
  {D.}~\bibnamefont {Ceresoli}}, \bibinfo {author} {\bibfnamefont
  {M.}~\bibnamefont {Cococcioni}}, \ and\ \bibinfo {author} {\bibnamefont {{\em
  et. al.}}},\ }\href {\doibase 10.1088/1361-648x/aa8f79} {\bibfield  {journal}
  {\bibinfo  {journal} {Journal of Physics: Condensed Matter}\ }\textbf
  {\bibinfo {volume} {29}},\ \bibinfo {pages} {465901} (\bibinfo {year}
  {2017})}\BibitemShut {NoStop}%
\bibitem [{\citenamefont {Bj\"orkman}(2011)}]{cif2cell}%
  \BibitemOpen
  \bibfield  {author} {\bibinfo {author} {\bibfnamefont {T.}~\bibnamefont
  {Bj\"orkman}},\ }\href {\doibase 10.1016/j.cpc.2011.01.013} {\bibfield
  {journal} {\bibinfo  {journal} {Computer Physics Communications}\ }\textbf
  {\bibinfo {volume} {182}},\ \bibinfo {pages} {1183 } (\bibinfo {year}
  {2011})}\BibitemShut {NoStop}%
\bibitem [{\citenamefont {Balogh}\ \emph {et~al.}(2006)\citenamefont {Balogh},
  \citenamefont {Rib\'{a}rik},\ and\ \citenamefont {Ung\'{a}r}}]{balogh:jap06}%
  \BibitemOpen
  \bibfield  {author} {\bibinfo {author} {\bibfnamefont {L.}~\bibnamefont
  {Balogh}}, \bibinfo {author} {\bibfnamefont {G.}~\bibnamefont {Rib\'{a}rik}},
  \ and\ \bibinfo {author} {\bibfnamefont {T.}~\bibnamefont {Ung\'{a}r}},\
  }\href {\doibase 10.1063/1.2216195} {\bibfield  {journal} {\bibinfo
  {journal} {Journal of Applied Physics}\ }\textbf {\bibinfo {volume} {100}},\
  \bibinfo {pages} {023512} (\bibinfo {year} {2006})}\BibitemShut {NoStop}%
\bibitem [{\citenamefont {Hart}\ \emph {et~al.}(1970)\citenamefont {Hart},
  \citenamefont {Aggarwal},\ and\ \citenamefont {Lax}}]{hart:prb70}%
  \BibitemOpen
  \bibfield  {author} {\bibinfo {author} {\bibfnamefont {T.~R.}\ \bibnamefont
  {Hart}}, \bibinfo {author} {\bibfnamefont {R.~L.}\ \bibnamefont {Aggarwal}},
  \ and\ \bibinfo {author} {\bibfnamefont {B.}~\bibnamefont {Lax}},\ }\href
  {\doibase 10.1103/PhysRevB.1.638} {\bibfield  {journal} {\bibinfo  {journal}
  {Phys. Rev. B}\ }\textbf {\bibinfo {volume} {1}},\ \bibinfo {pages} {638}
  (\bibinfo {year} {1970})}\BibitemShut {NoStop}%
\bibitem [{\citenamefont {Trivedi}\ \emph {et~al.}(1977)\citenamefont
  {Trivedi}, \citenamefont {Sharma},\ and\ \citenamefont
  {Kothari}}]{trivedi:jpcssp77}%
  \BibitemOpen
  \bibfield  {author} {\bibinfo {author} {\bibfnamefont {P.~C.}\ \bibnamefont
  {Trivedi}}, \bibinfo {author} {\bibfnamefont {H.~O.}\ \bibnamefont {Sharma}},
  \ and\ \bibinfo {author} {\bibfnamefont {L.~S.}\ \bibnamefont {Kothari}},\
  }\href {\doibase 10.1088/0022-3719/10/18/014} {\bibfield  {journal} {\bibinfo
   {journal} {Journal of Physics C: Solid State Physics}\ }\textbf {\bibinfo
  {volume} {10}},\ \bibinfo {pages} {3487} (\bibinfo {year}
  {1977})}\BibitemShut {NoStop}%
\bibitem [{\citenamefont {Men\'endez}\ and\ \citenamefont
  {Cardona}(1984)}]{menendez:prb84}%
  \BibitemOpen
  \bibfield  {author} {\bibinfo {author} {\bibfnamefont {J.}~\bibnamefont
  {Men\'endez}}\ and\ \bibinfo {author} {\bibfnamefont {M.}~\bibnamefont
  {Cardona}},\ }\href {\doibase 10.1103/PhysRevB.29.2051} {\bibfield  {journal}
  {\bibinfo  {journal} {Phys. Rev. B}\ }\textbf {\bibinfo {volume} {29}},\
  \bibinfo {pages} {2051} (\bibinfo {year} {1984})}\BibitemShut {NoStop}%
\bibitem [{\citenamefont {Debernardi}\ \emph {et~al.}(1995)\citenamefont
  {Debernardi}, \citenamefont {Baroni},\ and\ \citenamefont
  {Molinari}}]{debernardi:prl95}%
  \BibitemOpen
  \bibfield  {author} {\bibinfo {author} {\bibfnamefont {A.}~\bibnamefont
  {Debernardi}}, \bibinfo {author} {\bibfnamefont {S.}~\bibnamefont {Baroni}},
  \ and\ \bibinfo {author} {\bibfnamefont {E.}~\bibnamefont {Molinari}},\
  }\href {\doibase 10.1103/PhysRevLett.75.1819} {\bibfield  {journal} {\bibinfo
   {journal} {Phys. Rev. Lett.}\ }\textbf {\bibinfo {volume} {75}},\ \bibinfo
  {pages} {1819} (\bibinfo {year} {1995})}\BibitemShut {NoStop}%
\bibitem [{\citenamefont {Klotz}\ \emph {et~al.}(1997)\citenamefont {Klotz},
  \citenamefont {Besson}, \citenamefont {Braden}, \citenamefont {Karch},
  \citenamefont {Pavone}, \citenamefont {Strauch},\ and\ \citenamefont
  {Marshall}}]{klotz:prl97}%
  \BibitemOpen
  \bibfield  {author} {\bibinfo {author} {\bibfnamefont {S.}~\bibnamefont
  {Klotz}}, \bibinfo {author} {\bibfnamefont {J.~M.}\ \bibnamefont {Besson}},
  \bibinfo {author} {\bibfnamefont {M.}~\bibnamefont {Braden}}, \bibinfo
  {author} {\bibfnamefont {K.}~\bibnamefont {Karch}}, \bibinfo {author}
  {\bibfnamefont {P.}~\bibnamefont {Pavone}}, \bibinfo {author} {\bibfnamefont
  {D.}~\bibnamefont {Strauch}}, \ and\ \bibinfo {author} {\bibfnamefont
  {W.~G.}\ \bibnamefont {Marshall}},\ }\href {\doibase
  10.1103/PhysRevLett.79.1313} {\bibfield  {journal} {\bibinfo  {journal}
  {Phys. Rev. Lett.}\ }\textbf {\bibinfo {volume} {79}},\ \bibinfo {pages}
  {1313} (\bibinfo {year} {1997})}\BibitemShut {NoStop}%
\bibitem [{\citenamefont {Brazhkin}\ \emph {et~al.}(2000)\citenamefont
  {Brazhkin}, \citenamefont {Lyapin}, \citenamefont {Trojan}, \citenamefont
  {Voloshin}, \citenamefont {Lyapin},\ and\ \citenamefont
  {Mel'nik}}]{brazhkin:jetpl00}%
  \BibitemOpen
  \bibfield  {author} {\bibinfo {author} {\bibfnamefont {V.~V.}\ \bibnamefont
  {Brazhkin}}, \bibinfo {author} {\bibfnamefont {S.~G.}\ \bibnamefont
  {Lyapin}}, \bibinfo {author} {\bibfnamefont {I.~A.}\ \bibnamefont {Trojan}},
  \bibinfo {author} {\bibfnamefont {R.~N.}\ \bibnamefont {Voloshin}}, \bibinfo
  {author} {\bibfnamefont {A.~G.}\ \bibnamefont {Lyapin}}, \ and\ \bibinfo
  {author} {\bibfnamefont {N.~N.}\ \bibnamefont {Mel'nik}},\ }\href {\doibase
  10.1134/1.1320111} {\bibfield  {journal} {\bibinfo  {journal} {JETP Letters}\
  }\textbf {\bibinfo {volume} {72}},\ \bibinfo {pages} {195} (\bibinfo {year}
  {2000})}\BibitemShut {NoStop}%
\end{thebibliography}
%

\end{document}